\pdfoutput=1
\documentclass[aps,prd,nofootinbib,twocolumn,citeautoscript,showkeys,superscriptaddress]{revtex4-1} 
\usepackage{graphicx}  % needed for figures
\usepackage{dcolumn}   % needed for some tables
\usepackage{bm}        % for math
\usepackage{amssymb}   % for math
\usepackage{epsfig}
\usepackage{epstopdf}  
\usepackage{subfigure}
\usepackage{tensor}
\usepackage[utf8]{inputenc}
\usepackage{mdframed}
\usepackage{amsmath}
\usepackage{protosem}
\usepackage{wasysym}
\usepackage[bookmarks=false]{hyperref} 
\hypersetup{pdfstartview=FitH,pdfhighlight=/O,colorlinks=true}

\usepackage[toc]{appendix}
\usepackage{color,soul} 
\usepackage{datetime}

\newcommand{\labell}[1]{\label{#1}}

\newcommand{\bea}{\begin{eqnarray}}
\newcommand{\eea}{\end{eqnarray}}
\newcommand{\ba}{\begin{eqnarray}}
\newcommand{\ea}{\end{eqnarray}}

\newcommand{\beq}{\begin{equation}}
\newcommand{\eeq}{\end{equation}}
\newcommand{\beqa}{\begin{eqnarray}}
\newcommand{\eeqa}{\end{eqnarray}}
\newcommand{\beqar}{\begin{eqnarray*}}
\newcommand{\eeqar}{\end{eqnarray*}}

\newcommand{\ssc}{\scriptscriptstyle}
\newcommand{\eg}{{\it e.g.,}\ }
\newcommand{\ie}{{\it i.e.,}\ }

\newcommand{\req}[1]{(\ref{#1})} %{Eq.\thinspace(\ref{#1})}  

\begin{document}

\title{Universal black hole stability in four dimensions} 
\author{Pablo Bueno} 
\email{pablo@itf.fys.kuleuven.be}
\affiliation{Instituut voor Theoretische Fysica, KU Leuven,
Celestijnenlaan 200D, B-3001 Leuven, Belgium}
\author{Pablo A. Cano}
\email{pablo.cano@uam.es}
\affiliation{Instituto de F\'isica Te\'orica UAM/CSIC,
C/ Nicol\'as Cabrera, 13-15, C.U. Cantoblanco, 28049 Madrid, Spain}
\date{\today}
%\keywords{Modified theories of gravity, Black holes, Information paradox, Holography}
%\pacs{04.50.Kd, 04.50.-h, 04.60.-m}

\begin{abstract}
We show that four-dimensional black holes become stable below certain mass when the Einstein-Hilbert action is supplemented with higher-curvature terms. We prove this to be the case for an infinite family of ghost-free theories involving terms of arbitrarily high order in curvature. The new black holes, which are non-hairy generalizations of Schwarzschild's solution, present a universal thermodynamic behavior for general values of the higher-order couplings. In particular, small black holes have infinite lifetimes. When the evaporation process makes the semiclassical approximation break down (something that occurs after a time which is usually infinite for all practical purposes), the resulting object retains a huge entropy, in stark contrast with Schwarzschild's case.

\end{abstract}

\maketitle
%%%%%%%%%%%%%%%%%%%%%%%%%%%%%%%%%%%%%%%%%%%%%%%%%%%%%%%%%%%%%%%%%%%%%%
%%%%%%%%%%%%%%%%%%%%%%%%%%%%%%%%%%%%%%%%%%%%%%%%%%%%%%%%%%%%%%%%%%%%%%
%%%%%%%%%%%%%%%%%%%%%%%%%%%%%%%%%%%%%%%%%%%%%%%%%%%%%%%%%%%%%%%%%%%%%%
%%%%%%%%%%%%%%%%%%%%%%%%%%%%%%%%%%%%%%%%%%%%%%%%%%%%%%%%%%%%%%%%%%%%%%
\section{Introduction}
As proven by Hawking \cite{Hawking:1974sw}, a black hole with surface gravity $\kappa$ emits thermal radiation with a temperature $T_{\ssc \rm H}= \kappa/(2\pi)$. In the prototypical case of a Schwarzschild black hole \cite{Schwarzschild:1916uq} of initial mass $M_0$, the temperature increases as the black hole radiates, as a consequence of its negative specific heat. After a finite time of order $\sim M_0^3/M_{\rm \ssc P}^4$, where $M_{\rm \ssc P}$ is the Planck mass, the black hole evaporates down to an order-$M_{\rm \ssc P}$ object of order-one entropy. This suggests a violent ending for the evaporation process, and gives rise to the information paradox (see \cite{Harlow:2014yka,Chen:2014jwq} for recent reviews): a pure state collapsed to form a black hole would evolve into thermal radiation, in tension with the unitary evolution expected from quantum mechanics --- see \cite{Myers:1997qi}, though. Before the evaporation process has come to an end, one could hope that the black hole acts as an information reservoir (so that, \eg pure states are still pure when one does not trace out over the interior degrees of freedom). This is however very difficult to sustain, given the extremely low entropy of the Planck-mass object towards which the black hole evolves --- see also the discussion at the end of Section \ref{conclusions}.
%\footnote{And also, because one could in principle consider }

These considerations rely, in particular, on the somewhat unusual thermodynamic properties of the Schwarzschild black hole, which is expected to be the correct description of the exterior gravitational field of any static and spherically symmetric matter configuration for small enough curvatures. On the other hand, the effective gravitational action 
is expected to contain an infinite series of higher-derivative corrections modifying the usual Einstein-Hilbert term --- see \eg \cite{Nepomechie:1985us,Gross:1986iv,Gross:1986mw,tHooft:1974toh,Deser:1974cz}. One is then naturally led to wonder how these terms modify Schwarzschild's solution in higher-curvature situations and how such modifications might affect its thermodynamic behavior, as well as the evaporation process --- see \eg \cite{Myers:1988ze,Myers:1989kt,Callan:1988hs} for progress in this direction (mostly in higher-dimensions).

Unfortunately, constructing non-trivial extensions of the four-dimensional Schwarzschild black hole in higher-derivative gravities is a far from trivial task --- see \eg \cite{Lu:2015psa,Lu:2015cqa}.\footnote{In particular, our interest is on solutions which modify the Einstein gravity ones in a non-trivial way, and such that they reduce to the latter when the higher-derivative couplings are turned off. This discards, in particular: embeddings of Einstein gravity solutions into other theories, \eg \cite{delaCruzDombriz:2009et}; solutions to higher-order gravities which do not include an Einstein gravity term, \eg \cite{Riegert:1984zz,Klemm:1998kf}; solutions to theories which involve the fine-tuning of couplings of different orders in curvature \eg \cite{Love,Cai:2009ac}.}
For example, as opposed to the $D\geq 5$ cases, all the higher-order Lovelock invariants \cite{Lovelock1,Lovelock2} --- which due to their special properties do allow for such simple extensions, \eg \cite{Boulware:1985wk,Wheeler:1985qd,Myers:1988ze,Cai:2001dz,Dehghani:2009zzb,deBoer:2009gx} --- are  trivial in $D=4$.

Recently, the construction of \emph{Einsteinian cubic gravity} \cite{PabloPablo}, has triggered the construction of new four- (and higher-)dimensional generalizations of the Schwarzschild black hole \cite{Hennigar:2016gkm,PabloPablo2,Hennigar:2017ego,PabloPablo3,Ahmed:2017jod}. These solutions share the property of being fully determined by a single metric function, \ie they can be written in Schwarzschild-like coordinates as
\begin{equation}\label{fmetric}
	ds_f^2=-f(r)dt^2+f(r)^{-1} dr^2 + r^2 d\Omega_{(2)}^2\, ,
\end{equation}
a characteristic which turns out to be related to the absence of ghosts in the linearized spectrum of the theory \cite{Quasi,PabloPablo3}. Besides, the new black holes are non-hairy and fully characterized by their mass $M$, they describe the exterior field of spherical mass distributions \cite{PabloPablo3}, and reduce to Schwarzschild's solution when the corresponding higher-derivative couplings are set to zero.

In this paper, we will argue that these black holes are particular cases of an infinite family of four-dimensional solutions to a gravity theory involving terms of arbitrarily high order in curvature. As opposed to Schwarzschild's, the specific heat of the new black holes becomes positive below certain mass, which completely changes the evaporation process. 
In particular, the new small black holes have infinite lifetimes. Besides, following the evaporation process till the point in which the semiclassical approximation breaks down, one finds that a naturally enormous time is required and that the resulting object has a huge entropy, in flagrant contrast with the Schwarzschild black hole case --- see Table \ref{tbl} below for a summary.  In the small mass regime, the new (asymptotically flat) black holes satisfy the Smarr-type relation: $M=\frac{2}{3}T S$, which, intriguingly, coincides with the result corresponding to a three-dimensional CFT at finite temperature.
We argue that the above properties are universal for general values of the higher-order couplings.

\section{Higher-order corrections}\label{hoc}
The four-dimensional theories studied in \cite{PabloPablo,Hennigar:2016gkm,PabloPablo2,Hennigar:2017ego,PabloPablo3} and \cite{Ahmed:2017jod} are cubic and quartic representatives of an infinite family of higher-order gravities admitting simple extensions of the Schwarzschild solution whose thermodynamic properties can be easily accessed. The Lagrangian of this family can be written as
\begin{equation}\label{lal}
\mathcal{L}= \frac{1}{16\pi G}\left[R+\sum_{n=3}^{\infty}\frac{\lambda_n}{M_c^{2(n-1)}}\mathcal{R}_{(n)}\right]\, ,
\end{equation}
where $G=1/M_{\rm \ssc P}^2$ is the Newton constant, $M_c$ is some new energy scale, $\lambda_n$ are dimensionless couplings and $n$ is the order in curvature of each invariant $\mathcal{R}_{(n)}$, which will be formed by contractions of the metric and the Riemann tensor (but not its covariant derivatives).
%The characteristic feature of the $\mathcal{R}_{(n)}$ is the fact that \req{lal} allows for non-trivial extensions of Schwarzschild's solution characterized by a single function, \ie of the form \req{fmetric}, for general values of the $\lambda_n$. In particular, the metric function reduces to $f(r)=1-2GM/r$ when $\lambda_n=0$ for all $n$. 

The $\mathcal{R}_{(n)}$ are constructed using the following criterion. 
Let $L_f$ be the effective Lagrangian resulting from the evaluation of $\sqrt{|g|}\mathcal{L}$ on the ansatz \req{fmetric}, \ie
$
L_f\equiv \sqrt{|g|} \mathcal{L}\rvert_{g^{ab}=g_f^{ab}}. 
$
Then, we select the $\mathcal{R}_{(n)}$ in a way such that the Euler-Lagrange equation of $f(r)$ associated to $L_f$ vanishes identically, \ie for all $f(r)$. As shown in \cite{PabloPablo3}, any theory of the form \req{lal} fulfilling this requirement, satisfies the following properties: 1) it allows for solutions of the form \req{fmetric}, where $f(r)$ is determined by a second-order differential equation; 2) these solutions describe the exterior gravitational field of a static and spherically symmetric mass distribution; 3) the solutions are fully characterized by the total mass $M$; 4) the linearized equations on the vacuum are the same as Einstein gravity's (up to an overall constant), which in particular implies that the theory spectrum is free of ghost modes --- see appendix \ref{linearized}. 

Interestingly, for a given order $n$, there is a single possible way of modifying the $(n-1)$-order solution. In other words, once we find a particular density $\mathcal{R}_{n}^A$ contributing non-trivially to the equation determining $f(r)$, there does not exist a different density $\mathcal{R}_{n}^B$ which contributes in a different way to the same equation. There is however an important level of degeneracy on the $\mathcal{R}_{n}$ (growing with $n$), namely, there are several independent densities which give rise to the same equation (connected through the addition of densities which do not modify the equations of motion when evaluated on a general static and spherically symmetric ansatz). This is the case, for example, of the Einsteinian cubic gravity \cite{PabloPablo} and Generalized quasitopological gravity \cite{Hennigar:2017ego} densities in the cubic case.\footnote{In \cite{Hennigar:2017ego,Ahmed:2017jod}, careful analyses of these degeneracies were carried out for the cubic and quartic cases (not only in $D=4$, but in general dimensions).}

Our interest in this paper lies on the thermodynamic properties of the solutions, which are not affected by the particular representative $\mathcal{R}_{(n)}$ chosen at a given order. 
%Therefore, we will not put special emphasis on in the exact form of the $\mathcal{R}_{(n)}$ (as long as they exist and contribute in a unique possible way to the equation of $f$)
 %which are not affected by this issue (\ie such properties are the same independently of which representative among all possible invariants contributing in the same way to the $f(r)$ equation is chosen).
 Information about the explicit form of the $\mathcal{R}_{(n)}$ for general values of $n$ is provided in appendix \ref{holo}. For the sake of completeness, let us at least mention that particularly nice invariants for $n=3$ and $n=4$ are given, respectively, by: 
$
\mathcal{R}_{(3)}=-\frac{1}{6}(12\tensor{R}{_{a}^{b}_{c}^{d}}\tensor{R}{_{b}^{e}_{d}^{f}}\tensor{R}{_{e}^{a}_{f}^{c}} +R_{ab}^{cd} R_{cd}^{ef} R_{ef}^{ab} -12\tensor{R}{_{abcd}}\tensor{R}{^{ac}}\tensor{R}{^{bd}}+8R^{b}_{a} R_{b}^{c} R_{c}^{a})$ and $
\mathcal{R}_{(4)}=+\frac{2}{3}(	10 R^{abcd }R_{a \ c}^{\ e\ f}R_{e\ b}^{\ g\ h}R_{f g d h}+4R^{ab }R^{cd e f }R_{c\ e a}^{\ h}R_{d h f b}-14R^{abcd }R_{ab }^{\ \ ef }R_{c\ e}^{\ g \ h}R_{d g f h}-5R^{abcd}R_{a\ c}^{\ e\ f}R_{e\ f}^{\ g\ h}R_{b g  d h}).
$
These densities belong to the \emph{Einsteinian} class \cite{PabloPablo,Aspects}, \ie when considered in \req{lal}, the corresponding theories share the linearized spectrum of Einstein gravity not only in $D=4$, but also when considered in arbitrary higher dimensions.\footnote{It would be interesting  to find out whether one can always construct invariants $\mathcal{R}_{(n)}$ belonging to the \emph{Einsteinian} class for arbitrary $n$. }

\section{Black holes}\label{bhs}
 When evaluated in the ansatz \req{fmetric}, the field equations of \req{lal} reduce to the following (single) second-order differential equation for $f(r)$\footnote{In practice, we constructed explicitly the first three densities: $\mathcal{R}_{(3,4,5)}$, which allowed us to guess the pattern for general $n$ in \req{fequationn}. Then, we verified that the $n=6,7,8,9,10$ cases indeed agree with such pattern. Observe that the pattern is very simple, which strongly supports our claim that \req{fequationn} is also valid for general $n\geq 11$. We also checked that, in all cases, there is indeed a single possible way of modifying \req{fequationn} at a given order $n$. }   
\begin{widetext}
	\begin{align}\label{fequationn}
		&2GM-(1-f)r=-\sum_{n=3}^{\infty}\frac{\lambda_n}{M_c^{2(n-1)}}\left(\frac{f'}{r}\right)^{n-3}  \left[\frac{f'^3}{n}+\frac{(n-3)f+2}{(n-1)r}f'^2-\frac{2}{r^2}f(f-1)f'  -\frac{1}{r}f f''\left(f'r-2(f-1)\right)\right]
		\, ,
	\end{align}
\end{widetext}
where $M$ stands for the ADM mass of the solution \cite{Arnowitt:1960es,Arnowitt:1960zzc,Arnowitt:1961zz}. For $n=3$ and $n=4$, the above equation agrees with the ones found for Einsteinian cubic gravity \cite{PabloPablo2,Hennigar:2016gkm} and the quartic version \cite{Ahmed:2017jod} of Generalized quasitopological gravity \cite{Hennigar:2017ego}. The general version of this equation for arbitrary values of the cosmological constant and general horizon geometries can be found in appendix \ref{mastereq}.

Observe that all terms in the sum contribute to the equation in a very similar fashion, which will ultimately be responsible for the universal thermodynamic properties of the solutions. In order for \req{fmetric} to represent a black hole solution, we need to impose asymptotic flatness, as well as the existence of a regular horizon for some $r=r_h>0$. 

%\subsection{Asymptotic region}
In the asymptotic region, the Einstein gravity contribution dominates $f(r)$, and the higher-order corrections can be assumed to be small. These considerations\footnote{Further details on this kind of analysis can be found in \cite{PabloPablo2,Hennigar:2017ego,Ahmed:2017jod}.} lead to the following asymptotic expansion:
\begin{equation}\label{asymptexpn}
f(r)=1-\frac{2GM}{r}\left[1+\frac{18\lambda_3 (GM/r)^{5}}{(G M M_c)^4}+\cdots\right]+f_{\rm \ssc h}(r)\, ,
\end{equation}
where the dots stand for subleading contributions on $GM/r$ (including terms proportional to the higher-order couplings), and
\begin{equation}\label{homo}
f_{\rm \ssc h}(r)\sim A \exp \left(\frac{2r^{5/2}}{5G\sqrt{6\lambda_3 GM}}\right)+B \exp \left(\frac{-2r^{5/2}}{5G\sqrt{6\lambda_3 GM}}\right) \, .
\end{equation}
Hence, in order to get an asymptotically flat solution, we must set $A=0$, which is equivalent to fixing one of the two integration constants in \req{fequationn}.
Observe that in order for this conclusion to hold, we must demand the lowest-order non-vanishing $\lambda_n$ to be positive (we chose this to be $\lambda_3$ in \req{asymptexpn}). Otherwise, we would obtain non-decaying oscillating solutions which would not be asymptotically flat in general.
 %because otherwise we would obtain oscillating solutions which do not decay at infinity and the solution cannot be asymptotically flat in general. 
%If $\lambda_3=0$, then an analogous analysis shows that we must have $\lambda_4>0$ to get good asymptotics, and, in general, the first non-vanishing $\lambda_n$ must be positive in order to get a good asymptotic behavior.  
The second term in \req{homo} can be neglected for practical purposes, as the decaying exponential is extremely subleading in \req{asymptexpn}. 
%Finally, from \req{asymptexpn}, it is clear that $M$ is in fact the mass of the black hole.

%\subsection{Near-horizon region}
Imposing the existence of a regular event horizon fixes the remaining integration constant. In order to see this, it is convenient to perform a Taylor expansion at the horizon,
\begin{equation}
f(r)=4\pi (r-r_h)T+\sum_{n=2}^{\infty}a_n (r-r_h)^n,
\label{series}
\end{equation} 
where $T$ is the black hole temperature and $a_n=f^{(n)}(r_h)/n!$. Solving \req{fequationn} for the first two orders in $(r-r_h)$ gives rise to the following relations:
 \begin{align} \label{massnn}
 &2GM=r_h-\sum_{n=3}^{\infty}\frac{\lambda_n (4\pi T)^{n-1}}{M_c^{2n-2}r_h^{n-2}}\frac{(2n+(n-1)4\pi T r_h )}{ n(n-1)}\, , \\ 
 \label{temperaturenn}
 1=&4\pi Tr_h+\sum_{n=3}^{\infty}\frac{\lambda_n (4\pi T)^{n-1}}{M_c^{2n-2}r_h^{n-1}}\frac{(2n+(n-3)4\pi T r_h)}{n(n-1)} .
 \end{align}
These equations fix $r_h$ and $T$ in terms of the black hole mass $M$. Note that these relations are exact, as they are necessary conditions for having a smooth near-horizon geometry.
Depending on the values of the $\lambda_n$ and the mass, these equations can have one, several, or even no solutions at all. If there are several solutions, it means that various possible black holes with the same mass can exist. However, only one of the solutions will smoothly reduce to Schwarzschild in the limit $\lambda_n\rightarrow 0$ for all $n$. This is the one which should be regarded as physical. 
%On the other hand, for $M$ big enough there is always at least one solution, but it could happen that when $M$ is small no solution exists, meaning that there are no black holes below certain mass. This is indeed an intriguing possibility that we will not cover here. 
We will be mostly interested in the case in which there is a unique solution for any value of $M$, in whose case we can write  $T(M)$ and $r_h(M)$ without ambiguity. This happens, for example, if $\lambda_n\ge 0$ for all $n$,
%\footnote{This choice naturally depends on the choice and normalization of the invariants $\mathcal{R}_{(n)}$. The condition $\lambda_n\ge 0$ for all $n$ seems to be always enough to achieve the uniqueness condition for a suitable choice of $\mathcal{R}_n$, \eg like the one presented in appendix \ref{holo}.} 
which, at the same time ensures the solution to be well-behaved asymptotically.

\begin{figure}[tp]
		\includegraphics[scale=0.54]{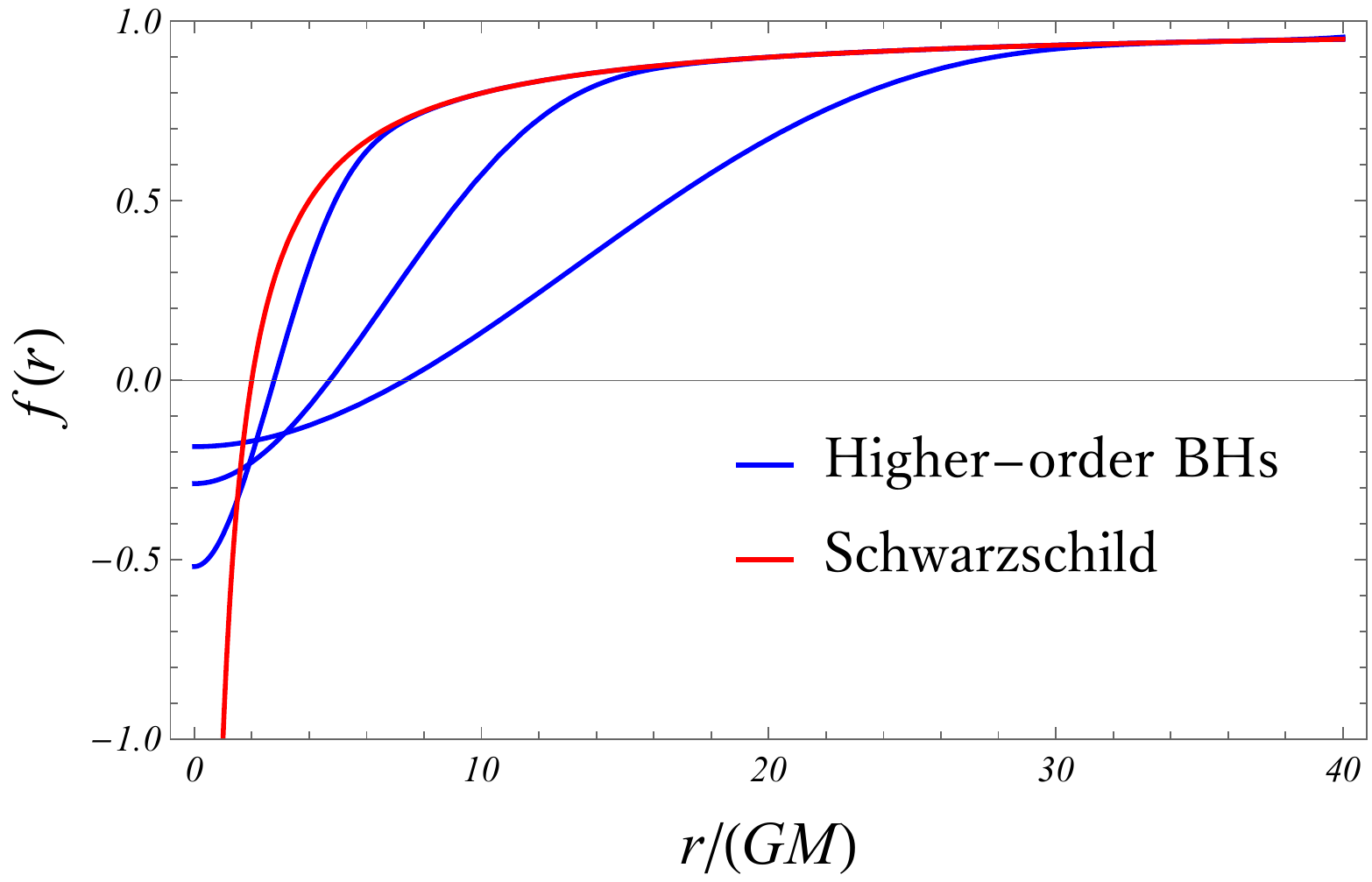}
		\caption{Metric function $f(r)$ for Schwarzschild's solution (red) and for the new higher-order black holes (blue), with $\lambda_3=\lambda_4=\lambda_5=\lambda_6=1$, $\lambda_{n>6}=0$ and $GMM_c=0.5\, ,0.2\, ,0.1$ respectively, (from left to right). }
		\label{fig108}
\end{figure}

 Once $T(M)$ and $r_h(M)$ have been determined, the $(r-r_h)^2$-order equation fixes $a_3$ as a function of $a_2=f''( r_h)/2$, the $(r-r_h)^3$ one fixes $a_4$ as a function of $a_3$ and $a_2$, and so on. In this process the only undetermined parameter is $a_2$, which means that the solution will be fully determined once we choose a value for it. Therefore, the family of solutions with a regular horizon is characterized by a single parameter, which must be carefully chosen so that the solution is asymptotically flat. 
% , {\rm i.e.}, we must glue a regular-horizoned solution with an asymptotically flat one. This can be done by choosing an appropriate value of $a_2$. Indeed, this value must be chosen with a high precision so that the growing exponential mode in \req{homo} is not excited. This is done by performing a numerical analysis, choosing a value for $a_2$ and computing the numerical solution. 
 In all cases studied, a numerical analysis shows that there is  a unique value of $a_2$ for which asymptotic flatness is achieved.  This means that there exists a unique asymptotically flat black hole, fully characterized by its mass $M$, which reduces to Schwarzschild's solution when the higher-order couplings are set to zero. The results presented in \cite{PabloPablo3} imply that, just like for  Schwarzschild in the Einstein gravity case, this will also represent the exterior field of a generic spherically symmetric matter distribution. 

In appendix \ref{numnum}, we provide a detailed discussion of the numerical construction of the solutions (which, unfortunately, do not seem accesible analitically). 
The resulting metric functions, $f(r)$, for a particular set of $\lambda_n$ and different values of $M_c$ are shown in  Fig. \ref{fig108}. Changing the $\lambda_n$ does not modify these curves qualitatively.

\section{Thermodynamics}

Making use of Wald's formula \cite{Wald:1993nt}, it is possible to compute the entropy of the solutions, which yields
\begin{align}\label{entropynn}
&S=\frac{\pi r_h^2}{G}\left[1-2\sum_{n=3}^{\infty}\frac{\lambda_n (4\pi T)^{n-1}}{M_c^{2n-2} r_h^{n-1}} \cdot \right. \\ \notag& \left. \cdot \left(\frac{2}{(n-2) 4\pi T r_h}+\frac{1}{n-1}\right)\right]+\frac{4\pi}{GM_c^2}\sum_{n=3}^{\infty}\frac{\lambda_n \chi^{n-2}}{(n-2)}\, ,
\end{align}
where $\chi$ is defined as
\begin{equation}
\sum_{n=3}^{\infty} \frac{2\lambda_n \chi^{n-1}}{(n-1)} \equiv 1\, .
\end{equation}
Using \req{entropynn}, we verify that the first law, $dM=T dS$, is satisfied. The last term in \req{entropynn} is a constant that ensures that $S\rightarrow 0$ as $M\rightarrow 0$, something that can always be done by adding a particular total derivative term to the original action \req{lal} (\eg an explicit Gauss-Bonnet contribution).

\begin{figure}[tp]
		\includegraphics[scale=0.542]{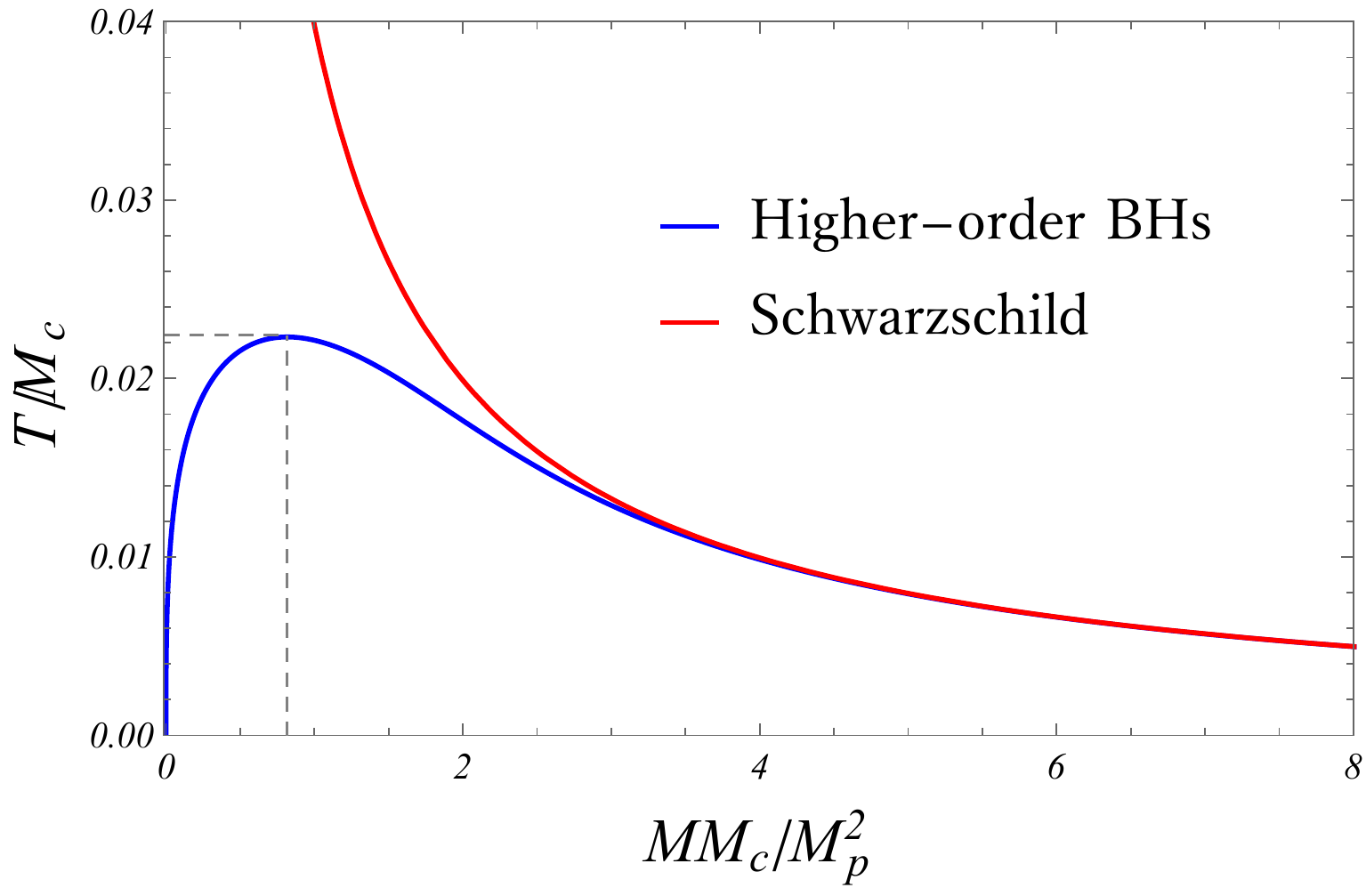}
	\caption{Black hole temperature as a function of the mass for Schwarzschild's solution (red) and for the higher-order black holes with $\lambda_3=\lambda_4=\lambda_5=\lambda_6=1$, $\lambda_{n>6}=0$ (note that the blue line is valid for any $M_c$). The higher-order black holes become stable below $M_{\rm \ssc max}\sim M_{\rm \ssc P}^2/M_c$. The shape of this curve is qualitatively the same for any other choice of couplings (except $\lambda_n=0$ for all $n$).}
	\label{fig1b}
\end{figure}

 As opposed to the  case of the Schwarzschild black hole, the temperature of the new solutions reaches a maximum for certain value of the mass, $M_{\rm \ssc max}\sim M_{\rm \ssc P}^2/M_c$. As shown in Fig. \ref{fig1b}, black holes with large enough masses behave like Schwarzschild's. However, when $M<M_{\rm \ssc max}$, the specific heat of the solutions becomes positive, which implies that small black holes behave as ordinary thermodynamic systems. Hence, for instance, they become colder as they radiate, and they tend to reach  equilibrium in the presence of a thermal environment. This behavior takes place as soon as one of the higher-order couplings is non-zero. Remarkably, turning on additional couplings does not modify this behavior qualitatively. All the possible black holes corresponding to the infitely many possible choices of $\lambda_n$ $(\geq 0)$ belong to the same universality class, the only exception being Schwarzshchild's black hole.

When the mass is much smaller than the corresponding maximum value, $M\ll M_{\rm \ssc max}$, the expressions for $r_h(M)$, $S(M)$ and $T(M)$ are approximately given by
\begin{align}\label{esi}
	r_h&=\left[\frac{M}{\zeta \chi ^3 M_c^2 M_{\rm \ssc P}^2} \right]^{1/3}\, ,\, \, S=6\pi \left[\frac{\zeta^{1/2}  M M_{\rm \ssc P} }{ M_c^2} \right]^{2/3} \, ,\\ \label{eso}
T&=\frac{1}{4\pi}\left[\frac{M M_c^4}{\zeta  M_{\rm \ssc P}^2} \right]^{1/3}\, , \text{  where} \quad \zeta\equiv \sum_{n=3}^{\infty} \frac{ \lambda_n\chi^{n-3}}{n}\, .
\end{align}
In this regime, the solutions satisfy the Smarr relation 
\begin{equation}\label{smmr}
M=\frac{2}{3}TS\,,
\end{equation}
which turns out to coincide with the result corresponding to a three-dimensional CFT at finite temperature (if we identify $M$ with the CFT energy density). Note that \req{smmr} is valid for arbitrary values of $\lambda_n$ as long as at least one of them is non-vanishing. This is a quite intriguing result. In fact, \req{smmr} coincides also with the result found for planar asymptotically Anti-de Sitter (AdS) black holes, as expected from the AdS/CFT correspondence \cite{Maldacena,Witten,Gubser}. In contrast, \req{smmr} holds here for the spherically symmetric and asymptotically flat black holes of the theory \req{lal} in the small mass limit. Note that \req{smmr} differs from Schwarzschild's analogous relation, which reads $M=2TS$ instead (observe that such relation is also valid for general $\lambda_n$ when $M\gg M_{\rm \ssc max}$).

\section{Black hole evaporation}
\begin{table*}[]
	\centering
	\begin{tabular}{c||c|c|c} 
		&  Semiclassical approximation breakdown mass& Time till  $M\sim M_{\rm  \ssc min}$     & Entropy when $M\sim M_{\rm  \ssc min}$\\
		\hline \hline
		Schwarzschild  & $M_{\rm  \ssc min}\sim M_{\rm \ssc P}$ & $\Delta t \sim M_0^3/M_{\ssc \rm P}^4 $  & $S \sim 1$ \\ \hline
		Higher-order BHs & $ M_{\rm  \ssc min}\sim  \sqrt{M_{\ssc \rm P}M_c}  $   & $\Delta t\sim M_{\rm \ssc P}^{7/2}/M_{c}^{9/2}$  & $S\sim M_{\rm \ssc P}/M_c$  %\\ 
	\end{tabular}
	\caption{We compare the result of the evaporation process for a Schwarzschild black hole to the one for the new higher-order solutions at the scale for which the semiclassical approximation stops making sense. When the minimum mass is reached, the Schwarzschild black hole has turned into a Planck-mass object of order-one entropy. For the new black holes, the entropy of the  resulting object is huge instead.  
	}
	\label{tbl}   
\end{table*}   
Let us now explore how the evaporation process of black holes gets affected by the special thermodynamic behavior of the new solutions in the small mass regime. The rate of mass-loss of a black hole in the vacuum can be computed using the Stefan-Boltzmann law, 
\begin{equation}
\frac{d M(t)}{dt}=-4\pi r_h^2 \sigma \cdot  T^4\, ,
\end{equation}
where $\sigma=\pi^2/60$. Using \req{esi}, we can easily integrate this expression for $M\ll M_{\rm \ssc max}$. The result is
\begin{equation}
	M(t)= \frac{M_0}{1+ t/t_{1/2}}\, ,\quad  \text{where} \quad t_{1/2}=\frac{3840\pi\chi^2\zeta^2 M_{\rm\ssc P}^4}{ M_0 M_c^4 }\, ,
\end{equation}
is the half-life of the black hole. Then, we observe that an infinite time is required for the black holes to evaporate completely.\footnote{A similar behavior was previously observed, for example in: \cite{Myers:1988ze,Myers:1989kt} for certain higher-dimensional Lovelock black holes; \cite{Callan:1988hs} for $D(>4)$-dimensional black holes in dilaton gravity modified with a stringy Riemann-squared term;   and  in \cite{Easson:2002tg} for $D=2$ dilaton gravity black holes.} Note also that the half-life is huge as long as $M_c\ll M_{\ssc \rm P}$.
This condition can be easily made  compatible with the requirement that the physics of macroscopic black holes does not get affected by the new couplings, which can be expressed as: $M_c \gg (G M_{\odot})^{-1}\sim 10^{-9}$ eV.
% a condition which is compatible with the requirement that the new couplings do not affect the physics of macroscopic black holes: $M_c \gg (G M_{\odot})^{-1}\sim 10^{-9}$ eV. 

The regime of validity of the semiclassical description  is characterized by the fact that the horizon radius is much greater than the corresponding Compton wavelength, \ie $r_h\gg \lambda_{\rm \ssc Compton}\sim 1/M$.
%In order for the semiclassical approximation to make sense, the horizon radius of our small black holes should be much greater than the associated Compton wavelength, $r_h\gg \lambda_{\rm \ssc Compton}\sim 1/M$. 
Similarly, the thermodynamic description is valid whenever $S\gg 1$, and breaks down  as $S\sim 1$. A Schwarzschild black hole reaches the Planck mass in a time $\Delta t\sim M_0^3/ M_{\rm \ssc P}^4$, after which both conditions are violated. In contrast, for the new black holes, the condition on the entropy is never violated as long as we choose $M_c\ll M_{\ssc \rm P}$. The semiclassical approximation stops making sense for a mass of order 
\begin{equation}
	M_{\rm  \ssc min}\sim \zeta^{1/4}\chi^{3/4} \sqrt{M_{\ssc \rm P}M_c}\,.
\end{equation}
Interestingly, when such a mass is reached, the entropy is still very large, namely
\begin{equation}
S_{\rm \ssc min} \sim \sqrt{\chi \zeta}\frac{M_{\rm \ssc P}}{M_c}\, .
\end{equation}
Starting with a black hole of initial mass $M_{\rm \ssc min } \ll M_0 \ll M_{\rm \ssc max }$, the minimum mass is reached in a time 
\begin{equation}
	\Delta t \sim \chi^{5/4}\zeta^{7/4} \frac{M_{\rm \ssc P}^{7/2}}{M_c^{9/2}}\, ,
\end{equation}
which, remarkably, does not depend on $M_0$ as long as $M_0 \gg M_{\rm \ssc min } $. We summarize these observations in Table \ref{tbl}.

It is illustrative to make a quantitative comparison with the Schwarzschild case. If we choose, say, $M_c\sim 1$ GeV, the temperature maximum would be at $M_{\rm \ssc max}\sim 10^{13}$ kg, and the semiclassical approximation break down mass would be 
$M_{\rm \ssc min}\sim 10^6$ TeV. At that point, the entropy would be of order $S_{\rm \ssc min}\sim 10^{19}$. In contrast, a Schwarzschild black hole of mass $M_0\sim 10^{6}$ kg, would evaporate down to the Planck mass in $\Delta t \sim 1$ minute, with a final entropy $S\sim 1$.
One of the new higher-order black holes with the same initial mass (or, in fact, with any mass such that $M_{\rm \ssc min } \ll M_0 \ll M_{\rm \ssc max } $), would need $\Delta t \sim 10^{25}$ times the age of the universe to reach $M_{\rm \ssc min}$. %For all practical purposes, the new black holes would live forever, and no conflict with non-unitary evolution would arise.  In addition, black holes with masses close to $M_{\rm \ssc min}$ would still posses a huge entropy, in glaring contrast with the Schwarzschild case.  \\
Hence, in contrast to the Schwarzschild case, the lifetime  of the new small black holes  (understood, in this case, as the period till the semiclassical approximation breakdown) is usually huge or, rather, infinite for all practical purposes.

Let us close this section by mentioning that one could of course consider the evaporation process of black holes with $M_0> M_{\rm \ssc max}$ as well. In that case, the process would be Schwarzschild-like till the black hole reaches the point for which the remaining mass is smaller than $M_{\rm \ssc max}$. After that, the discussion in this section would start applying. Of course, this implies that all black holes --- big or small --- belonging to the new family have infinite lifetimes.

\section{Conclusions}\label{conclusions}
 In this paper, we have presented an infinite family of four-dimensional higher-order gravities admitting non-hairy, static and spherically symmetric generalizations of the Schwarzschild black hole. The new theories, defined in \req{lal}, have the same linearized spectrum as Einstein gravity, which implies, in particular, that no ghost modes are propagated by metric perturbations on the vacuum. 
 
 The new solutions, which reduce to Schwarzschild's when all the higher-order couplings are set to zero, present a  thermodynamic behavior which notably differs from the one encountered in the Schwarzschild black hole. In particular, the specific heat of the black holes becomes positive below certain mass $M_{\rm \ssc max}\sim M_{\rm \ssc P}^2/M_c$, where $M_c$ is a new scale --- see Fig. \ref{fig1b}. This behavior is universal for all the possible choices of the higher-order couplings $\lambda_n \geq 0$ with the only exception of Schwarzschild's solution itself. 
 
 Even though the solutions cannot be constructed analytically --- see appendix \ref{numnum} for a detailed discussion of the numerical construction ---, the relevant thermodynamic magnitudes can be accessed exactly and, in particular, explicit expressions for the horizon radius, the temperature and the entropy as functions of the black hole mass $M$ can be written for small enough masses, $M \ll M_{\rm \ssc max}$ --- see equations \req{esi} and \req{eso}. In this regime, the black holes satisfy the Smarr relation $M=2/3\cdot TS$ for arbitrary values of the new couplings (as long as at least one of them is non-vanish) which, intriguingly, is analogous to the one found for a three-dimensional CFT at finite temperature.

The thermodynamic properties of the small higher-order black holes have dramatic consequences for the evaporation process. In particular, as opposed to the Schwarzschild case, the new black holes have infinite lifetimes, so one could speculate that they might act as \emph{eternal} information reservoirs evading possible unitarity violations. During the evaporation, the black holes require a (normally huge) time $\Delta t \sim M_{\rm \ssc P}^{7/2}/M_c^{9/2}$  to reach a mass $M_{\rm \ssc min}\sim \sqrt{M_{\rm \ssc P}M_c}$ for which the semiclassical description breaks down. At that point, the entropy of the resulting 
% \bl{\emph{remnant} is still huge, $S\sim M_{\rm \ssc P}/M_c$. This is in stark contrast with the Schwarzschild black hole case, which reaches the Planck mass with an order-one entropy, and hence fails to provide a reasonable possible account of the missing information. } 
object is still huge, $S\sim M_{\rm \ssc P}/M_c$. This is in stark contrast with the Schwarzschild black hole case, which reaches the Planck mass with an order-one entropy.
%, and hence flagrantly fails to provide a reasonable possible account of the missing information. 
One can only speculate on what would happen below $M_{\rm \ssc min}$. However, it is worth emphasizing that the tendency suggested by the evaporation process of the new black holes is very different from Schwarschild's: first, if we extrapolated the semiclassical approximation beyond $M_{\rm \ssc min}$, the Schwarzschild black hole would disappear soon after, while the new black holes would placidly keep on living forever; and second, at the semiclassical breakdown point, the issue of being left with an extremely low entropy object gets notably softened. 

At this point, it is convenient to stress that attempts to resolve the information paradox involving remnant-like objects \cite{Aharonov:1987tp} have been usually argued to be very difficult to digest \cite{Preskill:1992tc,Harvey:1992xk,Giddings:1994qt,Susskind:1995da}. This is, in particular, because one can consider the evaporation process of arbitrary initial black holes, which would imply that an arbitrarily large amount of information would need to be stored in the remnant, forcing such finite-energy object to have an infinite number of internal states (or, in other words, infinitely many ``species'' of remnants would need to exist, one for each possible initial state
collapsing to a black hole).   This raises the question of how an arbitrary amount of information could be carried within a Planck volume (or, more generally, the volume associated to the semiclassical breakdown scale) and naturally leads to a problematic infinite production rate of remnant pairs --- see \eg \cite{Giddings:1994qt}. 
While these remain outstanding issues,\footnote{See \cite{Giddings:1992kn,Callan:1992rs,Giddings:1992hh} for some possible ways out.} our results illustrate that the evaporation process can drastically change  when higher-order terms are considered in the gravitational effective action, suggesting new perspectives for the final fate of black holes or, at least, for the way they approach the regime in which the semiclassical description stops making sense.

Let us close with some possible future directions. Firstly, it would be convenient to understand how common the thermodynamic properties of the black holes presented here are among other (perhaps more general) higher-derivative corrections to the Einstein-Hilbert action. A different venue worth exploring would entail trying to find out whether some of the new higher-order invariants, $\mathcal{R}_{(n)}$, can actually arise as $\alpha^{\prime}$ corrections in concrete four-dimensional String theory effective actions. One would naively expect that, just like the  $\mathcal{R}_{(n)}$, such terms should not give rise to ghosts in the linearized spectrum\footnote{This might actually be more subtle in general. For example, in \cite{Prue} it was argued that String Theory can actually produce terms giving rise to ghost-like gravitons in the theory spectrum, but that those should be an artifact of the truncation performed in the effective action.}  so, at least from this perspective, there seems to be a chance that this is the case.

It would also be interesting to study other consequences derived from the possible existence of small stable black holes in four dimensions. For example, it has been often argued that microscopic black holes could be responsible for dark matter --- see \eg \cite{Chen:2002tu,Easson:2002tg,Afshordi:2003zb,Nozari:2005ah,Mureika:2012na}. While the existence of microscopic Schwarzschild black holes is not allowed due to their rapid evaporation, the stability of the new black holes seems to make them suitable for this purpose.

\vspace{-0.2cm}

\acknowledgements 
We are happy to thank Rob Myers and Tom\'as Ort\'in for useful discussions and comments. 
The work of PB was supported by a postdoctoral fellowship from
the Fund for Scientific Research - Flanders (FWO). The work of PAC was supported by a ``la Caixa Severo
Ochoa'' International pre-doctoral grant and in part by the Spanish Ministry of Science and
Education grants FPA2012-35043-C02-01 and FPA2015-66793-P and the Centro de Excelencia Severo
Ochoa Program grant SEV-2012-0249.

\onecolumngrid  
%\begin{center}  
%{\Large\bf Appendices} 
%\end{center} 
\appendix

\section{Explicit form of the higher-order invariants} \labell{holo}  
At each order in curvature, there do not exist two invariants $\mathcal{R}_{(n)}^A$ and $\mathcal{R}_{(n)}^{B}$ giving rise to two different non-trivial contributions to the equation determining the metric function, \req{fequationn}, compatible with the ansatz \req{fmetric}. There is a unique contribution, which is nothing but the one which appears inside the sum in the rhs of \req{fequationn}.  However, as we explained in the main text, the invariants $\mathcal{R}_{(n)}$ are  uniquely defined only up to the addition of terms which do not modify the equations of motion for a general static and spherically symmetric ansatz. For example, one can always modify $\mathcal{R}_{(n)}$ by the addition of the corresponding $n$-th order Lovelock density. In fact, there exists a considerable amount of freedom when choosing the set of $\mathcal{R}_{(n)}$ invariants for which \req{lal} admits solutions of the form \req{fmetric}. In the main text, we presented a particular choice of invariants for the $n=3,4$ cases. Those present the additional particularity of sharing the linearized spectrum of Einstein gravity (not only in four dimensions, but) for general $D$, while being defined in the same way  in arbitrary dimensions (\ie the relative couplings of each invariant do not depend on $D$, like in the case of Lovelock theories). The general procedure for constructing Einsteinian gravities, which was explained in \cite{PabloPablo,Aspects}, could be used to determine whether it is always possible to find examples of 
$\mathcal{R}_{(n)}$ belonging to this class for $n\geq 5$.

A simpler approach for constructing  high-order invariants consists in considering products of lower-order densities. In particular, we claim that $5$ densities suffice to construct examples of $\mathcal{R}_{(n)}$ at any order. These densities can be chosen to be
\begin{align}
R\, , \quad  Q_1\equiv R_{ab}R^{ab}\, , \quad  Q_2 \equiv R_{abcd}R^{abcd}\, , \quad C_1\equiv \tensor{R}{_{a}^{b}_{c}^{d}}\tensor{R}{_{b}^{e}_{d}^{f}}\tensor{R}{_{e}^{a}_{f}^{c}} \, , \quad C_2\equiv R_{ab}^{cd} R_{cd}^{ef} R_{ef}^{ab}\, .
\end{align}
Using the Ricci scalar, the two quadratic invariants, $Q_{1,2}$, and the two cubic  ones, $C_{1,2}$, we can construct $3$ independent invariants at quadratic order: $\{R^2,Q_1,Q_2 \}$; $5$ at cubic order $\{R^3, R Q_1, R Q_2,C_1,C_2\}$; $8$ at quartic order: $\{R^4, R^2 Q_1, R^2 Q_2, R C_1,R C_2, Q_1^2,Q_2^2, Q_1 Q_2 \}$; $12$ at quintic order, $\{R^5, R^3 Q_1, R^3 Q_2, R^2 C_1,R^2 C_2, R Q_1^2, R Q_2^2,R Q_1 Q_2, Q_1C_1,Q_1 C_2,Q_2 C_1,Q_2 C_2 \}$; 19 at sextic order; 25 at septic order; $36$ at octic order, $45$ at nonic order, and so on. Observe that this number grows considerably slower than the total number of independent invariants at each order in curvature \cite{0264-9381-9-5-003}. Using these, we have constructed the following explicit set  of invariants up to $n=10$:
 %\begin{align}
 %\mathcal{R}_{(3)}=&+R^3-32\tensor{R}{_{a}^{b}_{c}^{d}}\tensor{R}{_{b}^{e}_{d}^{f}}\tensor{R}{_{e}^{a}_{f}^{c}}+2 R_{ab}^{cd} R_{cd}^{ef} R_{ef}^{ab}-3 R R_{abcd}R^{abcd} \, , \\ \notag
 % \mathcal{R}_{(4)}=&+R^4+27 (R_{ab}R^{ab})^2+\frac{15}{4} {(R_{abcd}R^{abcd})}^2-32R \tensor{R}{_{a}^{b}_{c}^{d}}\tensor{R}{_{b}^{e}_{d}^{f}}\tensor{R}{_{e}^{a}_{f}^{c}}+2R  R_{ab}^{cd} R_{cd}^{ef} R_{ef}^{ab}-6R^2R_{ab}R^{ab}\\ \notag&-21  R_{ef}R^{ef}R_{abcd}R^{abcd} \, , \\ \notag
  %\mathcal{R}_{(5)}=&+R^5
  %\end{align}
 \begin{align} \label{r3}
 \mathcal{R}_{(3)}=&+\frac{1}{12}\left(R^3-32C_1+2 C_2-3 R Q_2\right) \, , \\
  \mathcal{R}_{(4)}=&-\frac{1}{288}\left(4R^4+108 Q_1^2+15 {Q_2}^2- 128R C_1+8R C_2-24R^2 Q_1 -84 Q_1 Q_2 \right)\, , \\ 
  \mathcal{R}_{(5)}=&+\frac{1}{2160}\big(5R^5+132 R Q_1^2+18 R Q_2^2-272 R^2 C_1+10 R^2 C_2-30 R^3 Q_1-102 R Q_1 Q_2+552 Q_1 C_1\\ \notag&-156 Q_2 C_1\big)\, ,\\
    \mathcal{R}_{(6)}=&-\frac{1}{11520}\big(4R^6+180 R^2 Q_1^2+33 R^2Q_2^2-384 R^3 C_1+8 R^3 C_2-24R^4Q_1-156R^2 Q_1 Q_2+768 R C_1 Q_1\\ \notag&-192 Q_1^3-12Q_2^3+1728C_1^2+64 C_1 C_2+144 Q_1^2Q_2\big)\, , \\
      \mathcal{R}_{(7)}=&+\frac{1}{967 680}\big(84R^7-504 R^5Q_1+168 R^4 C_2-5760 R^4 C_1+293 R^3 Q_2^2-1676 R^3 Q_1 Q_2+2180 R^3 Q_1^2\\ \notag &+11776 R^2 C_1 Q_1-4800 R C_1 C_2+51904 R C_1^2+208 Q_2^2 C_2-4064 Q_2^2C_1-832 Q_1Q_2C_2+16640 Q_1Q_2C_1\\ \notag &+832Q_1^2C_2-17024 Q_1^2C_1\big)\, ,
\\
      \label{r8}
           \mathcal{R}_{(8)}=&-\frac{1}{611 186 688}\big(18130 R^8-108780 R^6 Q_1+36260 R^5 C_2-1023592 R^5 C_1-19437 R^4 Q_2^2-31032 R^4 Q_1 Q_2\\ \notag&+139812 R^4 Q_1^2+6515280 R^3 Q_1 C_1-1881680 R^2 C_1 C_2+12416172 R Q_2^2 C_1-21222000 R Q_1 Q_2 C_1\\ \notag&-7220688 R Q_1^2 C_1-1073478 Q_2^4+6549648 Q_1 Q_2^3-13534416 Q_1^2 Q_2^2+9893184 Q_1^3 Q_2-642448 Q_2 C_1 C_2\\ \notag&+56702496 Q_2 C_1^2-870240 Q_1^4+1284896 Q_1 C_1 C_2-5812928 Q_1 C_1^2\big) \, , 
           \\
           \label{r9}
            \mathcal{R}_{(9)}=&+\frac{1}{13934592}\big(1820 R^9-29400 Q_1 R^7+6300 Q_2 R^7 -64896 C_1 R^6 + 4760 C_2 R^6+187596 Q1^2 R^5\\ \notag& - 90156 Q_1 Q_2 R^5 + 6999 Q_2^2 R^5+1285632 C_1 Q_1 R^4 - 43680 C_2 Q_1 R^4 - 640128 C_1 Q_2 R^4 + 12600 C_2 Q_2 R^4\\ \notag& -1208768 C_1^2 R^3 - 55872 C_1 C_2 R^3 + 2240 C_2^2 R^3 - 
 767856 Q_1^3 R^3 + 855996 Q_1^2 Q_2 R^3 - 338064 Q_1 Q_2^2 R^3 \\ \notag&+ 51015 Q_2^3 R^3 -5208960 C_1 Q_1^2 R^2 + 100512 C_2 Q_1^2 R^2 + 3737856 C_1 Q_1 Q_2 R^2 - 66912 C_2 Q_1 Q_2 R^2 \\ \notag&- 332448 C_1 Q_2^2 R^2 + 8328 C_2 Q_2^2 R^2 -705792 C_1^2 Q_1 R + 137472 C_1 C_2 Q_1 R + 1596672 Q_1^4 R \\ \notag&+ 5192256 C_1^2 Q_2 R - 131136 C_1 C_2 Q_2 R - 2717568 Q_1^3 Q_2 R + 1580544 Q_1^2 Q_2^2 R - 340704 Q_1 Q_2^3 R\\ \notag& + 15120 Q_2^4 R+23224320 C_1^3 - 591360 C_1^2 C_2 - 53760 C_1 C_2^2 + 5117952 C_1 Q_1^3 + 111360 C_2 Q_1^3 \\ \notag&- 6398592 C_1 Q_1^2 Q_2 - 58560 C_2 Q_1^2 Q_2 + 2888448 C_1 Q_1 Q_2^2 - 24960 C_2 Q_1 Q_2^2 - 484320 C_1 Q_2^3 + 13200 C_2 Q_2^3\big)\, , 
\end{align}\begin{align}
\label{r10}
   \mathcal{R}_{(10)}=&-\frac{1}{39813120}\big(2100 R^{10}-6300 R^8 Q_2-12600 R^8 Q_1+8400 R^7 C_2+3840 R^7 C_1 -46435 R^6 Q_2^2\\ \notag&+210940 R^6 Q_1 Q_2-160 540 R^6 Q_1^2-12600 R^5 Q_2C_2-266880 R^5 C_1 Q_2 -25200 R^5 Q_1 C_2-270080 R^5 Q_1 C_1\\ \notag&+11625 R^4 Q_2^3 +448260 R^4 Q_1 Q_2^2-1856940 R^4 Q_1^2 Q_2+1827840 R^4 Q_1^3 +8400 R^4 C_2^2-291520 R^4 C_1 C_2\\ \notag&-194368 R^4 C_1^2-87670 R^3 C_2 Q_2^2+2265984 R^3 Q_2^2 C_1+325480 R^3 Q_1 Q_2 C_2-6631296 R^3 Q_1 Q_2 C_1\\ \notag&-300280 R^3 Q_1^2 C_2+8883456 R^3 Q_1^2 C_1-201348R^2 Q_2^4+1731744 R^2 Q_1 Q_2^3-6767712 R^2 Q_2^2 Q_1^2\\ \notag&+12733056 R^2 Q_1^3 Q_2-1037760 R^2 Q_2 C_1 C_2+34126272 R^2 Q_2 C_1^2-9027648 R^2 Q_1^4+2309120 R^2 Q_1 C_1 C_2\\ \notag&-46106624 R^2 Q_1 C_1^2+51600 R Q_2^3 C_2-3286368 R Q_2^3 C_1+62400 R Q_1 Q_2^2 C_2+13847808 R Q_1 Q_2^2 C_1\\ \notag&-868800 R Q_1^2 Q_2 C_2-9561216 R Q_1^2 Q_2 C_1+1075200 R Q_1^3 C_2-9977856 R Q_1^3 C_1-598400 R C_1 C_2^2\\ \notag&+2492800 R C_1^2 C_2+113305600 R C_1^3+126000 Q_2^5-705600 Q_1 Q_2^4-604800 Q_1^2 Q_2^3+10483200 Q_1^3 Q_2^2\\ \notag&+10400 Q_2^2 C_2^2-1041600 Q_2^2C_1 C_2-3990272 Q_2^2C_1^2-22176000Q_1^4 Q_2-41600Q_2 Q_1 C_2^2+5260800Q_1 Q_2 C_1 C_2\\ \notag&+11138048 Q_1 Q_2 C_1^2+14515200 Q_1^5+41600 Q_1^2 C_2^2-6355200 Q_1^2 C_1 C_2+19489792Q_1^2 C_1^2\big)\, .
  \end{align}
  
Yet  a different approach would entail repeating the construction performed in sections \ref{hoc} and \ref{bhs} in arbitrary dimensions. The idea would be then constructing sets of invariants (with dimension-dependent coefficients) admitting solutions satisfying  $g_{tt}g_{rr}=-1$, according to the criteria explained in section \ref{hoc}. The resulting densities would belong to the \emph{Generalized quasitopological gravity} class in the terminology of \cite{Hennigar:2017ego,Ahmed:2017jod} (the cubic and quartic cases were developed in those papers). Note that the densities \req{r3}-\req{r10} (as well as the Einsteinian ones presented in the main text for $n=3,4$) will be related to such densities through the addition of invariants which do not contribute to the equations of motion when evaluated on a general static and spherically symmetric ansatz.
  
Ideally, one would like to find a more systematic way of constructing sets of invariants $\mathcal{R}_{(n)}$, or even a closed-form expression from which the densities can be automatically read off. The techniques developed in \cite{Deser:2005pc} could be useful for this purpose --- see also the discussion in \cite{Cisterna:2017umf}.

\section{Numerical construction of the solutions}\label{numnum}
In this appendix we explain the numerical procedure which allows us to construct the black hole solutions analyzed in the main text. The same procedure has been previously used \eg in \cite{PabloPablo2,Hennigar:2017ego,PabloPablo3,Ahmed:2017jod}.
First, we observe that \req{fequationn} is a \emph{stiff} differential equation. This is because the terms involving derivatives of $f(r)$ appear as corrections, and, in particular, the coefficient that multiplies $f''$ is usually small. Thus, the numerical resolution is problematic in terms of stability and we need to use, at least, \emph{A-stable} methods. In our calculations, we used implicit Runge-Kutta methods, but, of course, other methods with larger stability regions can be used as well. Once we have an appropriate numerical method, \req{fequationn} can be solved by imposing the boundary conditions explained in the text.

In principle, we should start the solution at the horizon $r=r_h$, where we know $r_h$ and $a_1=f'(r_h)$ in terms of the mass and the solution is specified once we choose a value of $a_2=f''(r_h)/2$, as explained in the main text. In practice, the numerical method cannot be started at $r_h$, since at that point the equation is singular. Instead, we start the solution for some other value, $r_h+\epsilon$, very close to the horizon ($\epsilon \ll 1$). Then, the second-order Taylor polynomial of $f$ around the horizon can be used to compute $f(r_h+\epsilon)$ and $f'(r_h+\epsilon)$. This yields $f(r_h+\epsilon)=a_1\epsilon+a_2\epsilon^2$,  $f'(r_h+\epsilon)=a_1+2\epsilon a_2$. The numerical resolution can then be started at $r_h+\epsilon$ by using these initial conditions, where the only free parameter is $a_2$. %Of course, $\epsilon$ has to be as small as possible in order to obtain a reliable solution. 
Then, $a_2$ is chosen by imposing the solution to be asymptotically flat. From the asymptotic expansion analysis in section \ref{bhs}, we know that there exists a family of solutions which are exponentially growing when $r\rightarrow\infty$. Almost any choice of $a_2$ will excite this mode and the solution will not be asymptotically flat. Indeed, we expect that there is a unique choice of $a_2$ for which the solution is asymptotically flat. 

In order to find $a_2$, we use the shooting method to glue the numeric solution with the asymptotic expansion \req{asymptexpn}. The idea is the following: we first fix a value $r_{\infty}$ sufficiently large, for which the asymptotic expansion $f_{\rm asympt.}(r_{\infty})$ is a good approximation. Then we choose a value for $a_2$ and we compute numerically the solution up to $r_{\infty}$, which would yield a value $f_{\rm numeric}(r_{\infty}; a_2)$. The appropriate value of $a_2$ is such that it glues both solutions: $f_{\rm numeric}(r_{\infty}; a_2)=f_{\rm asympt.}(r_{\infty})$. In all the cases analyzed, there is  a unique value of $a_2$ for which this happens, and it must be chosen with great precision.  In practice, it is difficult to extend the numeric solution to very large $r$, since the equation becomes more and more stiff, but it is always easy to compute the numeric solution up to a value for which it overlaps with the asymptotic expansion. 

Once $a_2$ has been determined, the interior solution $r<r_h$ can also be computed by starting the numeric method at $r_h-\epsilon$. In this case, the numerical resolution offers no problems.

\begin{table}[ht]
	\begin{center}
		\centering
		\begin{tabular}{|l|l|}
			\hline
			$G M M_c$&\multicolumn{1}{c|}{$a_2/M_c^2$ } \\ \hline
			1& - 0.11391418211405159081 \\ \hline
			0.5&  - 0.05469962891492144178 \\ \hline
			0.2&  +0.10107506480043540696\\ \hline
			0.1 & +0.20446663339860675846\\ \hline
		\end{tabular}
		\caption{Initial condition $a_2$ for several values of $M$ for $\lambda_{\geq 4}= 0$, $\lambda_3=1$. The equation is solved setting $\epsilon=10^{-6} r_h$.}
			\label{my-label}
	\end{center}
\end{table}

Let us be even more explicit. In order to do so, we restrict ourselves to the case $\lambda_{\geq 4}= 0$. This leaves us with a single correction to the Einstein-Hilbert action in \req{lal}, which can be taken to be the Einsteinian cubic gravity term \cite{PabloPablo}. Since there is a single density, we can set $\lambda_3=1$ and absorbe the coupling in the definition of $M_c$. For a given mass $M$, we compute $r_h(M)$ and $a_1(M)$. Equivalently, $a_2$ will also be a function of the mass, although we do not know it explicitly.  Applying the previous method to various values of $M$, we find the values of $a_2$ shown in Table \ref{my-label}. These can be used to construct the numerical solution with $\epsilon=10^{-6} r_h$. In Fig. \ref{figa2}, we show $a_2(M)$ constructed from the interpolation of discrete values. In the limit $M>>(G M_c)^{-1}$, we recover the Schwarzschild value $a_2=-(2 G M)^{-2}$. On the other hand, its is possible to prove that
\begin{equation}
\lim_{M\rightarrow 0} a_2=\lim_{M\rightarrow 0}\frac{a_1}{2 r_h}=\frac{\chi}{2} M_c^2\, ,
\end{equation}
which is in fact valid for arbitrary values of the $\lambda_n$.
Finding an explicit expression for $a_2(M)$ (or at least a more efficient way of computing it for different values of the couplings) would be of interest, in particular in order to perform additional studies of the solutions presented here \eg in the contexts of holography or gravitational phenomenology, \eg along the lines of \cite{Brigante:2007nu,deBoer:2009pn,Camanho:2009vw,Buchel:2009tt,Cai:2009zv,Camanho:2009hu,Buchel:2009sk,Myers:2010jv,Quasi} and \cite{Holdom:2016nek}, respectively.

\begin{figure}[ht!]
	\centering 
	\includegraphics[scale=0.56]{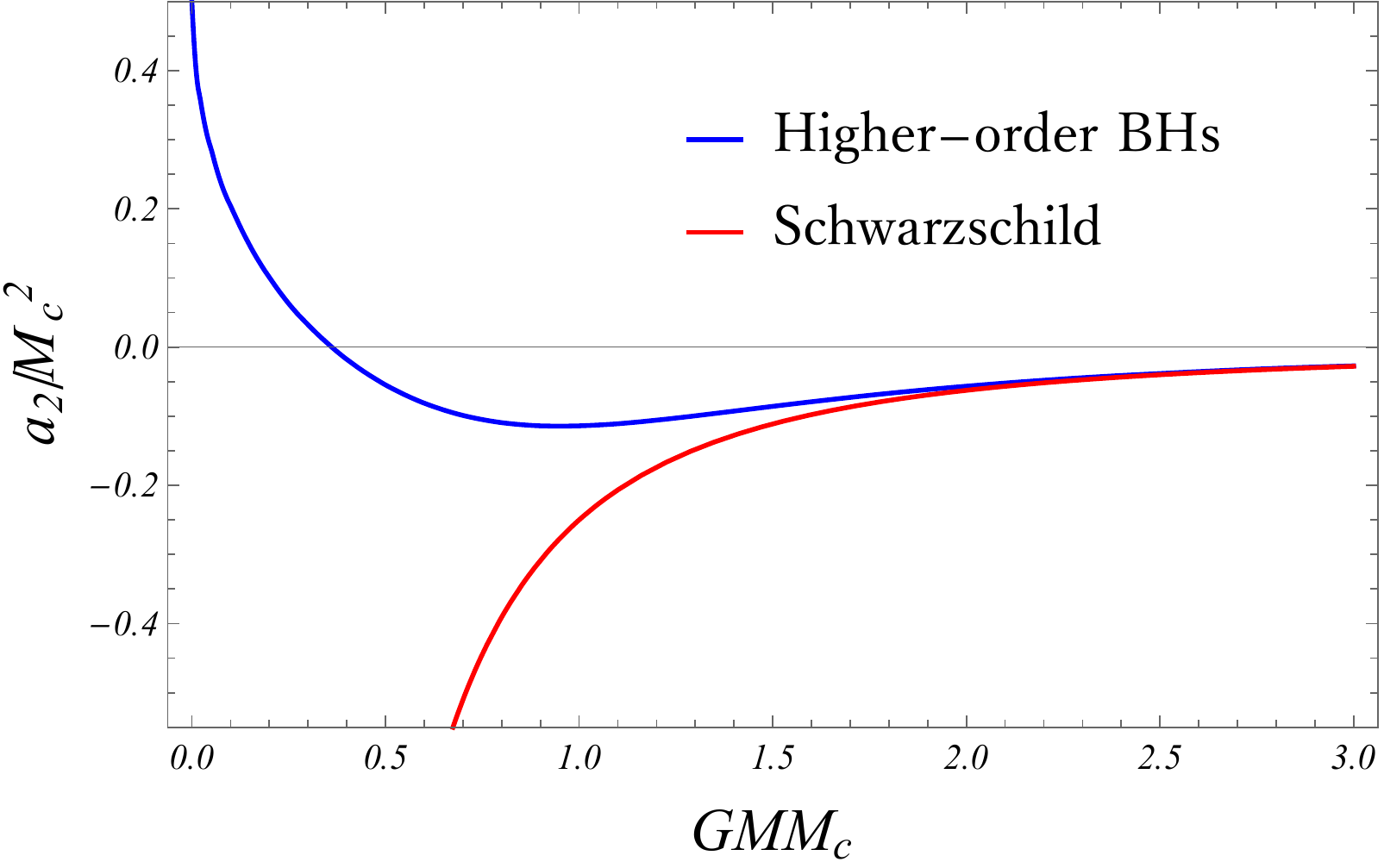}
	\caption{We plot $a_2/M_c^2$ as a function of the mass (interpolation) for  $\lambda_{\geq 4}= 0$, $\lambda_3=1$ and for the Schwarzschild solution, $a_2=-(2 G M)^{-2}$ (red).} 
	\labell{figa2}
\end{figure}

\section{(A)dS asymptotes and general horizon geometries}\label{mastereq}
In the main text we have focused on static, spherically symmetric and asymptotically flat solutions. Our results can be easily extended to (A)dS asymptotes as well as to planar or hyperbolic horizon geometries. In particular \req{lal} can be generalized to
\begin{equation}
\mathcal{L}= \frac{1}{16\pi G}\left[R-2\Lambda_0+\frac{\lambda_2}{M_c^2}\mathcal{X}_4+\sum_{n=3}^{\infty}\frac{\lambda_n}{M_c^{2(n-1)}}\mathcal{R}_{(n)}\right]\, ,
\end{equation}
where we have now made explicit the topological Gauss-Bonnet term $\mathcal{X}_4=R^2-4R_{ab}R^{ab}+R_{abcd}R^{abcd}$, which we will use to impose the entropy to vanish as $M\rightarrow 0$. We can search for solutions of the form
\begin{equation}
ds^2=-f(r)dt^2+\frac{dr^2}{f(r)}+r^2d\Sigma_{(k)}^2\, ,
\label{kmetric2}
\end{equation}
where $d\Sigma_{(k)}^2$ is the metric of a 2-dimensional maximally symmetric space of curvature $k=1,0,-1$, \ie, the metric of the unit sphere, flat space or hyperbolic space. The  generalized version of the equation which determines the metric function $f(r)$ then reads (note that $M$ is no longer the mass in the planar and hyperbolic cases):
\begin{align}
&-(f-k)r-\sum_{n=3}^{\infty}\frac{\lambda_n}{M_c^{2n-2}}\left(\frac{f'}{r}\right)^{n-3}\Bigg[\frac{f'^3}{n}+\frac{(n-3)f+2k}{(n-1)r}f'^2-\frac{2}{r^2}f(f-k)f'-\frac{1}{r}f f''\left(f'r-2(f-k)\right)\Bigg]\\ \notag &=2GM+\frac{1}{3}\Lambda_0 r^3\, .
\label{kfequationn2}
\end{align}
The generalized versions of equations \req{massnn} and \req{temperaturenn} read
%By performing a Taylor expansion around the horizon we can obtain the relations among $r_h$, $T$ and $M$:
\begin{align}
2GM-k r_h+\frac{1}{3}\Lambda_0 r_h^3&=-4\pi T\sum_{n=3}^{\infty}\frac{\lambda_n}{M_c^{2n-2}}\left(\frac{4\pi T}{r_h}\right)^{n-2}\left(\frac{4\pi T r_h}{n}+\frac{2k}{n-1}\right)\, ,\\
k-r_h^2\Lambda_0&=+4\pi Tr_h+\sum_{n=3}^{\infty}\frac{\lambda_n}{M_c^{2n-2}}\left(\frac{4\pi T}{r_h}\right)^{n-1}\frac{2nk+(n-3)4\pi T r_h}{n(n-1)}\, .
\end{align}
Wald's entropy reads in turn
\begin{equation}
\mathsf{S}=\frac{\pi r_h^2}{G}\left[1-2\sum_{n=3}^{\infty}\frac{\lambda_n}{M_c^{2n-2}}\left(\frac{4\pi T}{r_h}\right)^{n-1}\left(\frac{2k}{(n-2) 4\pi T r_h}+\frac{1}{n-1}\right)\right]+\frac{2\pi k\lambda_2}{G M_c^2}\, .
\label{entropynn2}
\end{equation}
%We can compute explicitly the thermodynamics in the limit $M<<M_{\rm max}$, corresponding to small stable black holes. We obtain the following relations, 
%\begin{equation}
%r_h=\left[\frac{M}{\zeta \chi ^3 M_c^2 M_{\rm \ssc P}^2} \right]^{1/3},\quad T=\frac{1}{4\pi}\left[\frac{M M_c^4}{\zeta  M_{\rm \ssc P}^2} \right]^{1/3},\quad \mathsf{S}=6\pi \left[\frac{\zeta^{1/2}  M M_{\rm \ssc P} }{ M_c^2} \right]^{2/3}+\frac{2\pi M_p^2}{M_c^2}\left(\lambda_2-2\sum_{n=3}^{\infty}\frac{\lambda_n}{n-2}\chi^{n-2}\right).
%\end{equation}
%In these expressions $\chi$ and $\zeta$ are constants which are determined through the equations
%\begin{equation}
%\sum_{n=3}^{\infty}\frac{2\lambda_n}{n-1}\chi^{n-1}=1,\quad \zeta=\frac{\Lambda_0}{3M_c^2\chi^3}+\sum_{n=3}^{\infty}\frac{\lambda_n}{n}\chi^{n-3}.
%\end{equation}

\section{Linearized equations}\label{linearized}
We can obtain the embedding equation of a maximally symmetric background of curvature $\Lambda$ by plugging $f(r)=k-\Lambda r^2$ in \req{kfequationn2} with $M=0$. Then we get the equation
\begin{equation}
\Lambda+M_c^2\sum_{n=3}^{\infty}\frac{\lambda_n (-1)^n}{n(n-1)}\left(\frac{2\Lambda}{M_c^2}\right)^n=\frac{1}{3}\Lambda_0\, ,
\label{Lambda}
\end{equation}
which determines the possible vacua of the theory. For any of them, we know that the theory only propagates a massless graviton, due to the results in \cite{PabloPablo3}.\footnote{We have verified this explicitly for all densities in appendix \ref{holo}.} Thus, the linearized equations satisfied by a metric perturbation $h_{ab}$ over a maximally symmetric background $\bar{g}_{ab}$, read simply
\begin{equation}
G_{ab}^L=8\pi G_{\rm eff} T_{ab}\, ,
\end{equation}
where $G_{ab}^L$ is the linearized Einstein tensor, 
\begin{align}
G_{ab}^L&=R_{ab}^L-\frac{1}{2}\bar g_{ab}R^L-(D-1)\Lambda h_{ab}\, ,\\
R_{ab}^L&=\bar\nabla_{(a|}\bar\nabla_{c}h^{c}_{\ |b)}-\frac{1}{2}\bar\Box h_{ab}-\frac{1}{2}\bar\nabla_{a}\bar\nabla_{b}h+D \Lambda h_{ab}-\Lambda h \bar g_{ab}\, , \\ \label{ricciscalar}
R^{L}&=\bar\nabla^{a}\bar\nabla^{b}h_{ab}-\bar \Box h-(D-1)\Lambda h\, ,
\end{align}
and where the effective gravitational constant $G_{\rm eff}$ reads
\begin{equation}
G_{\rm eff}=\frac{G}{1+2\sum_{n=3}^{\infty}\frac{\lambda_n (-1)^n}{n-1}\left(\frac{2\Lambda}{M_c^2}\right)^{n-1}}\, .
\end{equation}
Observe that, for theories with Einstein-like spectrum, this can be generically obtained by using the fact that $G/G_{\rm eff}$ is the slope of the lhs of \req{Lambda} evaluated on the background \cite{PabloPablo,Aspects}.
Finally, in the transverse gauge, $\bar \nabla_{a}h^{ab} =\bar \nabla^{b}h$, we can write
\begin{equation}
-\bar\Box h_{ab}=16\pi G_{\rm eff} T_{ab}\, .
\end{equation}

\renewcommand{\leftmark}{\MakeUppercase{Bibliography}}
\phantomsection
\bibliography{Gravities}
\label{biblio}

\end{document}